\documentclass[reprint,amsmath,amssymb,aps]{revtex4-1}
\usepackage{graphicx}
\usepackage{dcolumn}
\usepackage{bm}
\usepackage[utf8]{inputenc}
\begin{document}
\preprint{APS/123-QED}
\title{The role of memory in transition from direct to indirect reciprocity}
\author{Jurica Hižak}
\email{jurica.hizak@unin.hr}
\author{Lovorka Gotal Dmitrović}%
 \email{lgotaldmitrovic@unin.hr}
\affiliation{%
 University North\\
}%
\author{Mirko Čubrilo}
 \email{mirko.cubrilo@foi.hr}
\affiliation{
 Faculty of Organization and Informatics \\
 University of Zagreb
}%
\date{\today}

\begin{abstract}
It has been suggested that direct reciprocity operates well within small groups of people where it would be hard to get away with cheating one another but no research has been done yet to show how exactly the mechanism of direct reciprocity fails to operate as the group size increases. Unlike previous models that have neglected the role of memory, our model takes into account the memory capacity of the agents as well as the cost of having such memory. We have shown that the optimal memory capacity for handling the exploiters grows with the group size in a similar way as the relative size of the neocortex grows with the group size of the primates as it was found by Robin Dunbar. The time required for reaching the relative fitness of the defectors increases rapidly with the group size which points to the conclusion that there is an upper group size limit over which the mechanism of direct reciprocity is insufficient to maintain the cooperation. 
\end{abstract}
\maketitle
\section{\label{sec:level1}Introduction}
When it comes to the evolution of cooperation, the tit-for-tat (TFT) strategy is a sort of a celebrity among the strategies and constitutes one of the pillars, which ensures that Darwin's theory of natural selection does not collapse under the evidence of the altruistic behavior among non-relatives in nature \cite{axelrod1981,axelrod2006the,Milinski1987}. As computer simulations show, the TFT strategy wins the ultimately defective strategy always-defect  (ALLD) and gives rather convincing arguments in favor of Trivers' theory of direct reciprocity (DR). This was proven within classical framework of Iterated Prisoner’s Dilemma (IPD) as well as within stochastic IPD with reactive strategies \cite{Nowak1992}. Theoretically, if the probability of the encounter is large enough and assuming that TFT agents have a perfect memory and enough time available, the relative fitness value of the TFT strategy sooner or later exceeds the relative fitness of the ALLD strategy, regardless of the size of the population \cite{broom2013game-theoretical}. However, in nature there is no infinite memory, nor the infinitely large group; the memory capacity is limited, which probably limits the group size as well. There is empirical evidence which supports that idea – a British anthropologist Robin Dunbar has discovered that the relative size of the neocortex is correlated with the group size – more developed primates live in larger packs \cite{Dunbar1992}. Dunbar's discovery reveals that the size of the brain is adjusted to the size of the group because the brain is not a “computer for the orientation in the physical environment", but primarily "a computer for the orientation in the social environment" \cite{Dunbar1998,DavidBarrett2013,Adolphs2009}. Information that brain has to deal with is about making social bonding, alliances, detecting free-riders etc. 

In order to cope with free-riders, the individuals i.e. cooperative agents ought to memorize 3 types of information: “what”,“who” and “when”. For example a particular memory-2 agent might memorize defection (“what”) performed by agent no.3 ("who") in two previous interactions (“when”). Although there are studies that examine the role of memory in the evolution of cooperation, most of them seem to be concerned primarily with the limitations regarding the number of previous interactions which agents can take into account \cite{Stewart2016,Baek2016,Hilbe2017}.  It is often overlooked that agents do not interact with a single partner, but are embedded in a large social network. Despite the apparent simplicity of the memory-one strategies (such as TFT) which takes into account only the co-player’s previous move, tracking the reciprocal obligations within large group may place a computationally significant burden on memory systems \cite{hertwig2013simple}. However, the classic computational model of DR does not include the memory limitations regarding the number of co-players nor explains the transition from direct to indirect reciprocity.

The role of gossip in the evolution of cooperation has been seriously considered since Richard D. Alexander suggested that individuals will tend to help those who helped others \cite{alexander1987the}. This kind of mechanism is known as Indirect Reciprocity (IR) and numerous models have been proposed to explain ways in which it may operate. The most famous and probably the most influential model of IR was designed by Martin Nowak and Karl Sigmund in 1998. To each player they assigned a value called an \textit{image score} which could be considered as a reputation of a player. The image score varies according to the players kindness –it rises whenever the player helps others, and falls whenever the player defects. They arranged the game so that each player can take part in many rounds, but not with the same partner twice. However, the players could discriminate other players according to the information they were given about their reputation \cite{Nowak1998}. Their model has shown that nature prefers cooperators, but still there were some criticisms concerning the model.  The key weakness of the image scoring is that it punishes the player who defects on a defector, therefore it fails to distinguish justified from unjustified defections \cite{Panchanathan2003}.  Secondly, their model does not include DR, since the players cannot re-meet. It was shown that IR through image scoring becomes unstable with respect to DR as the probability of re-meeting increases \cite{Roberts2008}. To resolve the first issue considering the problem of justified/unjustified defection, more sophisticated strategies were proposed- those that can take into account the past of the co-player but also the past of the co-player’s co-players, and so on. It was shown that discrimination based on the concept of "justified defection" can lead to a stability if players have the same reputation in the eyes of all members of their population. But if players have different views about the reputation of others, then errors in perception can undermine cooperation \cite{Nowak2005,MartinezVaquero2013}. Gossip might be a way of achieving consensus and there are researchers who believe that IR should be based on gossip indeed \cite{Sommerfeld2007,Giardini2011,Giardini2016}. Giardini and Vilone have shown that a large quantity of gossip is necessary to support cooperation in Public Goods Game, and that group structure can reduce the effects of errors in transmission \cite{Giardini2016}. 

All those issues have motivated us to design a realistic tit-for-tat agents with a limited memory capacity who can move around and re-meet, who can gradually forget their experience and who can exchange the experiences of their own interactions when they meet. In Sec.II of this paper we define the model of the "oblivious TFT" strategy (OTFT). We introduce: (1) the memory capacity (2) the half-life of the memory decay and (3) the energy cost of having the memory. In Sec.III we describe the simulation experiments on OTFT agents with different memory capacities. Only when convinced that the results of the simulations fit Dunbar’s findings, we shall introduce the GossipTFT players with ability to exchange their own experiences (Sec.IV) and provide the evidence that the gossip players can cope with the defectors efficiently even in the very large groups.   
\section{\label{sec:level1}Model of the oblivious TFT}
In this paper, we apply the well-known model, namely the PD game to the multi-agent NetLogo modeling environment. In our model, agents move around randomly on the surface without borders which may be considered as the finite toroidal space. Each player is surrounded by his own private space that can be violated by the other individuals in which case they play one round of the PD game. After such an encounter they turn away from each other and move on. The probability of the encounter depends, of course, on the density of the agents, while the outcome of the game depends on their strategies and their history. At the beginning of the simulation each TFT player is given a list as large as the population with the elements which represent the surrounding agents. The list can be considered as memory, since the boolean value of each element depends on the previous experience. The TFT player starts with the list filled with default values false but if betrayed by someone, the list is updated in a way that the value false at the position numerically assigned to that particular individual changes into the value \textit{true}. For example, if the player experiences a betrayal from the third agent, then the history list is updated as follows:
\[[false,false,true,false,\ldots false].\]
Next time, when the TFT player re-meets the agent no.3, he plays defect as well. Therefore, the player who uses the TFT strategy can correct his behavior towards the defector and improve the overall score. If we run the simulation with ALLD and TFT in the same proportion of players, the relative fitness of the ALLD players is higher in the first stage of the simulation, however as time goes by the relative fitness of the TFT strategy increases and finally exceeds the relative fitness of the ALLD strategy.

The TFT players we have described so far have perfect and infinitely large memory – their  history lists to store the information about the other players can stretch to infinity and they never forget who is who in the population, which is obviously unrealistic. To make our model more realistic we have introduced the memory capacity – a number that limits the storage of booleans \textit{true}. When the number of boolean values \textit{true} reaches the upper limit $m$, the player cannot change the list elements any more – he is allowed to play the PD game with another exploiter but he cannot store the information \textit{true}. In other words, if he detects $m$ free-riders, he will store exactly $m$   elements \textit{true} and hence the list will be locked.
However, it is meaningless and unrealistic to store such information for an infinite amount of time. According to "use it or lose it" principle, we forget what we rarely see, and we are strengthening the memory of what we meet often. The phrase “use it or lose it” usually refers to the relationship between exercise and muscle mass - the muscle cells become larger after exposure to physical stimulation, but many studies have shown that similar process takes place in brain. For example, new neurons in the adult hippocampus that are stimulated by new and challenging learning experiences are more likely to survive and become incorporated into brain circuits. On the other hand most of the cells will die in one week unless engaged in some kind of learning experience \cite{Shors2012}. Therefore, it is reasonable to implement some kind of mechanism that would gradually remove old information. We usually see forgetfulness as a flaw, but in the combination with a limited memory, forgetfulness may be advantageous - memory decay cleans the list and frees the space to store some new information. The sources of forgetting are a matter of intense debate - researchers disagree about whether memories fade as a function of the mere passage of time (decay theory) or as a function of interfering succeeding events (interference theory) \cite{Barrouillet2011}. Over the years there have been sharp critiques of decay, questioning whether it plays any role at all \cite{Jonides2008}. However, this explanation has strong intuitive appeal, therefore we have designed the mechanism that works similarly to radioactive decay – each element true of the memory list has a certain likelihood of falling into the default state \textit{false}. A random number from 1 to 100 is chosen, in each time unit (which is called a tick in the NetLogo), and if the number is less than the percentage of probability $p$ the element true transforms into the false. Thus, the rate of memory decay can be expressed with half-life equation.

\begin{figure}[h!]
  \caption{The memory decay can be expressed either by the probability of an element to transform, either by the half-life. }
  \centering
  \includegraphics[width=0.5\textwidth]{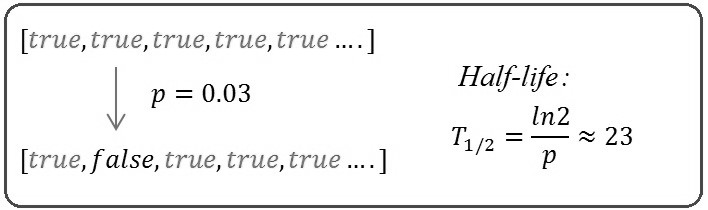}
\end{figure}
The third important feature of the “oblivious TFT” (OTFT) is the cost that player is obliged to pay for the memory. In the world of nature, having a brain is not just an advantage, but also a burden from the energetic perspective\cite{Dukas1999,Burns2010}. Having a memory is costly, thus we have assumed that the cost is proportional to the memory capacity with the constant of proportionality $k$, which we shall refer to simply as k-value. In each encounter, the fraction of energy \(\Delta E =k\times m\) is subtracted from the player’s payoff.
\section{\label{sec:level1}Restrictions of direct reciprocity}
Having designed those three important features of the OTFT, we have run the simulations with different k-values and half-lives, measuring the relative fitness of the ALLD and OTFT. Each population consisted of ALLD and OTFT with the same proportion of agents. The most representative results were obtained with \(T _\frac{1}{2} = 800 \) ticks and \(k = 5 \cdot 10^{-3}\). Using these values the simulation showed an obvious difference between players using different memory capacities.  Given a fixed population size, the OTFT players with larger memory capacity performed better, but only to a certain extent because with the increase of capacity the energy cost becomes so high that their relative fitness cannot reach the relative fitness of ALLD. 
The time needed to reach the relative fitness of ALLD does not decrease linearly, rather shows an interesting U-shape, which depends on half-life and k-values. 
\begin{figure}[h!]
  \caption{The time needed to reach the relative fitness of the ALLD strategy for different memory capacities and different group sizes (40-140 individuals). }
  \centering
  \includegraphics[width=0.5\textwidth]{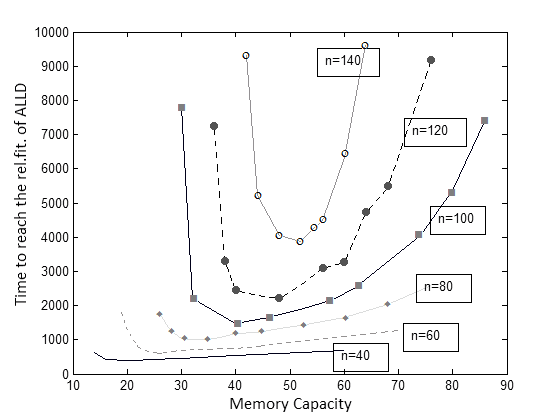}
\end{figure}
The stochasticity of the IPD is the result of the random walk of the agents which influences the performance, thus it is impossible to obtain the same exact result. We averaged the measured times and after more than 200 simulations conducted on different population sizes we have obtained curves with minima which shift the position nonlinearly and increases drastically along the vertical axis. The minimal time needed to take over the dominance increases asymptotically with the size of the group. Clearly, the asymptote represents the invisible border that cannot be crossed – if the group size exceeds approx.160 individuals, it is impossible for the cooperators to perform better, no matter how large is their memory capacity. 
\begin{figure}[h!]
  \caption{The minimal time needed to reach the relative fitness of the ALLD rises asymptotically with the group size.  }
  \centering
  \includegraphics[width=0.5\textwidth]{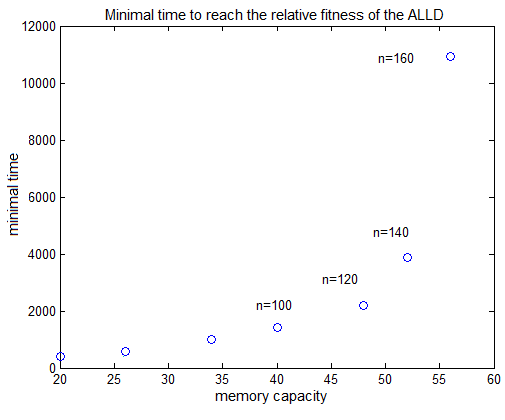}
\end{figure}
It must be emphasized that the probability of the encounter was conserved since we have programmed the simulation in such way that the size of the world was adjusted i.e. proportional to the number of agents.The minimal time corresponds to the optimal memory capacity suitable for a given group size. Amazingly, the optimal capacity is correlated to the group size with the regression coefficient r=0.9943. Such a strong correlation undoubtedly shows that our model faithfully represents what was suggested and empirically confirmed in the research of primates.
\begin{figure}[h!]
  \caption{The optimal memory capacity for a given group size increases linearly with the size of the population. }
  \centering
  \includegraphics[width=0.5\textwidth]{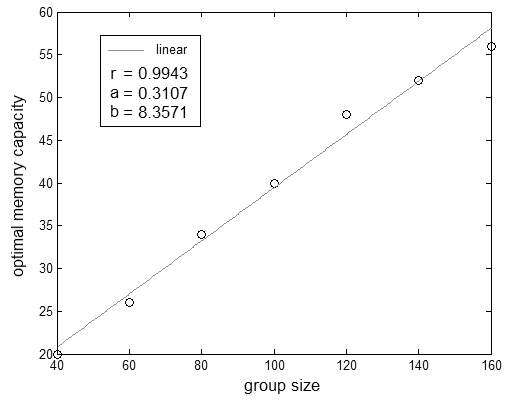}
\end{figure}
\section{\label{sec:level1}	The Gossip TFT}
Learning via direct encounters involves risk of being cheated. A less dangerous and therefore more optimal learning method is witnessing the experiences of others. This method usually called vicarious learning has been observed in a variety of animals from drosophila to mice \cite{Mobbs2015}. In humans, vicarious learning is transmitted through multiple channels including imitation, conversation and gossip. 

In order to cross the group size border we have designed a special type of TFT players who can exchange the information stored in their history lists. Unlike other authors, we didn’t try to build the reputation, but rather to use the instruments provided so far in the framework of DR. Instead of using the image score, we decided to rely on the history lists of players engaged in the encounter. The new kind of player that we named GossipTFT is able to alter the boolean values of its list when they get in touch with their kind. The procedure goes as follows: If two GossipTFT players meet both of them update their lists according to the simple logical rule:  \textit{true} OR \textit{false}  with the same position on the lists results in \textit{true}  to both of them. The rule may be thought of as precaution which is easy to be justified from the evolutionary perspective –“If he had bad experience with that guy, maybe it’s better to be cautious.” The asymmetry in consequences between false negatives (which can result in being hurt) and false alarms has often led the evolution in the direction of setting the fear threshold quite low (so that even light stimuli are actually interpreted as dangerous) \cite{Adolphs2013}. 
\begin{figure}[h!]
  \caption{If GossipTFT players meet both of them update their lists according to the simple logical rule:  true OR false with the same position on the lists results in true to both of them. }
  \centering
  \includegraphics[width=0.5\textwidth]{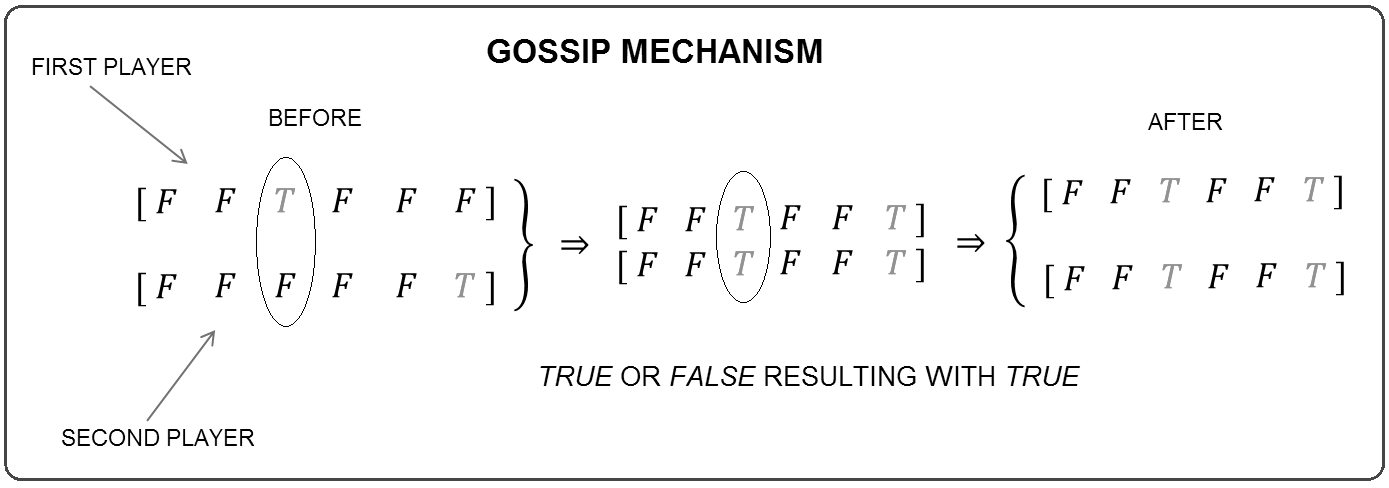}
\end{figure}

The GossipTFT players possess the same features, as the OTFT players; limited memory which decays over time and the energy cost of having the memory. Clearly, the cost of having the ability to gossip and to remember must be higher than the ability just to remember.  That is why we decided to use k-values  
\(k > 5 \cdot 10^{-3}\)
When pushed into the arena to fight versus the ALLD players, the GossipTFT players show an amazing performance. To make it harder for the GossipTFT, we have chosen a k-value \(k = 8 \cdot 10^{-3}\), but even then their speed is almost incomparable;   in the small groups they can reach the relative fitness of the ALLD strategy four times faster than the OTFT strategy. In the large groups they can make it forty times faster. 
\begin{figure}[h!]
  \caption{The performance of the OTFT players (with \(k = 5 \cdot 10^{-3}\)) compared to the GossipTFT players (with \(k = 8 \cdot 10^{-3}\)). }
  \centering
  \includegraphics[width=0.5\textwidth]{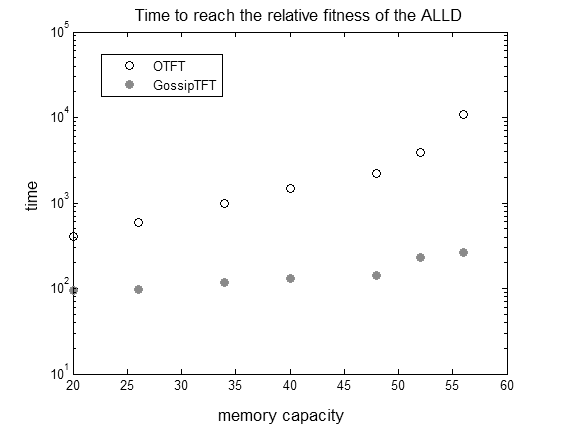}
\end{figure}
\section{\label{sec:level1}	Discussion}
It has been suggested that direct reciprocity operates well within small groups of people where it would be hard to get away with cheating one another \cite{nowak2011supercooperators} but no research has been done yet to show \textit{how exactly} the mechanism of direct reciprocity fails to operate as the group size increases. Unlike previous models that have neglected the role of memory, our model takes into account the cognitive abilities of the individuals as well as the cost of having those abilities.  As it was shown, we have modified the TFT strategy and introduced: (1) the memory capacity (2) the half-life of the memory decay and (3) the energy cost of having the memory. Using the NetLogo simulations on such “oblivious TFT” (OTFT) we have shown that the optimal memory capacity for handling the exploiters grows with the group size precisely as Dunbar empirically found. Moreover, we have shown that the time required for reaching the relative fitness of the defectors increases asymptotically with the population size which points to the conclusion that there is an upper group size limit at approximately 160 individuals over which the mechanism of direct reciprocity is insufficient to maintain the cooperation. It seems that above the upper limit it is necessary to allow the horizontal transmission of the information i.e. rumors about the free-riders present in the population. In order to allow such transmission without conceptual modifications of the starting model we have equipped our agents with the ability to exchange their own experiences with one another. Such agents, namely GossipTFT players, do not operate by discriminating the individuals according to their reputation, but rather by collecting the experiences of other GossipTFT players. The knowledge collected from others as well as the knowledge gained from the direct encounters may be fading since they forget the information over time, but it seems that these cognitive shortcomings do not reduce their performance. The GossipTFT players can cope with the defectors efficiently much faster than OTFT players even in the large groups and even when the energy cost of gossiping is significantly higher than the cost of just memorizing. The benefit of the information transmission through the population obviously exceeds the price paid for being able to communicate. 

It is believed that the demand for social cooperation via indirect reciprocity has propelled the evolution of human language\cite{nowak2011supercooperators} and this paper supports that belief – the stability of a large group is possible, as we have shown, only if there exist a communication which enables the denouncement of the exploiters. Thanks to language, humans are able to overcome the limitations of direct observations and can exchange information about each other, thus isolating defectors and selecting cooperative partners \cite{Giardini2011}. One might argue that in a real world, gossip is unreliable and cannot lead to the consensus about the kindness of the players, but still –in larger societies, especially in the structured societies there may be some other mechanisms of reputation besides the \textit{image score}. The image of an individual may be determined by the performance of its subgroup as well.  If such information, namely the \textit{group score} is available then the agents do not need to have perfect knowledge about others’ agent individual histories. On the contrary, as it was shown recently, for large populations only a tiny proportion of image scoring is sufficient to maintain the cooperation \cite{Nax2015}.

However, there are many features that have not been discussed nor included in the model, for example, the reproduction of the agents. The rise of the relative fitness over time, results in the increase of the probability of having the offspring, but the passage of time has relative meaning depending on the life span and the birth rate of the population. So far we have proven that time needed to reach the relative fitness of the defectors goes to infinity while the group grows, but we will not understand completely the transition from DR to IR, as long as we neglect the reproduction. We can only speculate that the critical group size might be even smaller when individuals reproduce since the arrival of the newborn members demands even a larger memory capacity. The implementation of the mechanism of reproduction into the small group brings in a whole series of problems since random drift can strongly affect evolution, but this work remains yet to be done. \\
\textbf{We thank Vinko Zlatić, PhD, Institute Rudjer Bošković, for his support and help.} 
\bibliographystyle{unsrt}
\bibliography{bibliography.bib}

\end{document}